\documentclass[twocolumn,showpacs,preprintnumbers,amsmath,amssymb]{revtex4}

\usepackage{graphicx}
\usepackage{dcolumn}
\usepackage{bm}

\begin{document}


\title{Characterization of multiqubit pure-state entanglement}
\author{Zeqian Chen}
\email{zqchen@wipm.ac.cn}
\affiliation{%
State Key Laboratory of Magnetic Resonance and Atomic and
Molecular Physics and United Laboratory of Mathematical Physics,
Wuhan Institute of Physics and Mathematics, Chinese Academy of
Sciences, 30 West District, Xiao-Hong-Shan, P.O.Box 71010, Wuhan
430071, China}
\author{Quanhua Xu}
\email{quanhua.xu@math.univ-fcomte.fr}
\affiliation{%
Laboratoire de Math\'{e}matique, Universit\'{e} de Franche-Comt\'{e},
16, Route de Gray, 25030 Besan\c{c}on Cedex, France}%

\date{\today}

\begin{abstract}
A necessary and sufficient entanglement criterion based on
variances of Mermin-Klyshko's Bell operators is proved for
multiqubit pure states. Contrary to Bell's inequalities, entangled
pure states strictly satisfy a quadratic inequality but product
ones can attain the equality under some local unitary
transformations, which can be obtained by solving a quadratic
maximum problem. This presents a characterization of multiqubit
pure-state entanglement.
\end{abstract}

\maketitle

Entanglement plays a crucial role in quantum communication,
cryptograph, and computation \cite{B-G-N}. The characterization of
entanglement is one of the fundamental problems in quantum
information theory \cite{B02}. Although there are some
characterizations of entanglement for bipartite systems
\cite{G,P-H} and many entanglement witnesses for testing
multipartite entanglement \cite{ME}, among which the most well
known one is the violation of Bell's inequalities \cite{MK} that
was originally designed to rule out various kinds of local hidden
variable theories based on Einstein, Podolsky, and Rosen's (EPR's)
notion of local realism \cite{EPR,B}, simple necessary and
sufficient conditions are not known for multipartite entanglement,
even in the simplest case of multiqubits \cite{CWKO}. In
particular, \.{Z}ukowski {\it et al} \cite{ZBLW} show that there
are some generalized GHZ states of $n>2$ qubits that do not
violate any Bell inequality for two dichotomic observables per
qubit, including the inequality involving the mean values of
Mermin-Klyshko's (MK) Bell operators. Since any pure entangled
state of two qubits violates a Clauser-Horne-Shimony-Holt (CHSH)
\cite{CHSH} inequality (Gisin's theorem \cite{G}), that is, the
CHSH inequality characterizes the pure-state entanglement of two
qubits, Scarani and Gisin \cite{SG} wrote that ``this analysis
suggests that MK inequalities, and more generally the family of
Bell's inequalities with two observables per qubit, may not be the
`natural' generalizations of the CHSH inequality to more than two
qubits".

However, in this article we find a necessary and sufficient
entanglement criterion based on variances of MK's Bell operators
for multiqubit pure states. This can be regarded as a
generalization of Gisin's theorem from two qubits to $n$ qubits,
although it cannot be straightforwardly generalized to all
multiqubit systems for the case of the Bell inequalities of two
dichotomic observables per site involving the mean values of
observables. Our criterion presents an operational
characterization of multiqubit pure-state entanglement, because it
is based on inequalities for variances of observables. Contrary to
the Bell inequalities, entangled pure states strictly satisfy the
inequality but product ones can attain the equality under some
local unitary transformations, which can be obtained by solving a
quadratic maximum problem. Our result sheds considerable new light
on relationships between multipartite entanglement and Bell's
operators.

Let us give a brief review of MK Bell operators and the associated
Mermin-Klyshko inequalities. The MK Bell operators of $n$ qubits
is defined recursively ($n \geq 2$). Let
$\vec{a}_j\vec{\sigma}_j,\vec{a}'_j\vec{\sigma}_j$ denote spin
observables on the $j$-th qubit, $j=1,...,n,$ where all
$\vec{a}_j, \vec{a}'_j$ are unit vectors in $\mathbb{R}^3$ and
$\vec{\sigma}_j = (\sigma^j_x, \sigma^j_y, \sigma^j_z)$ is the
Pauli matrices on the $j$-th qubit. Denote by ${\cal B}_1 =
\vec{a}_1 \vec{\sigma}_1$ and ${\cal B}'_1 = \vec{a}'_1
\vec{\sigma}_1.$ Define\begin{equation}{\cal B}_n = {\cal B}_{n-1}
\otimes \frac{1}{2}(\vec{a}_n\vec{\sigma}_n +
\vec{a}'_n\vec{\sigma}_n) + {\cal B}'_{n-1} \otimes \frac{1}{2}(
\vec{a}_n\vec{\sigma}_n - \vec{a}'_n\vec{\sigma}_n
),\end{equation}where ${\cal B}'_n$ denotes the same expression
${\cal B}_n$ but with all the $\vec{a}_j$ and $\vec{a}'_j$
exchanged. ${\cal B}_n$ is called the MK Bell operator of $n$
qubits. Assuming ``local realism" \cite{EPR,B}, one concludes the
Mermin-Klyshko inequality of $n$ qubits as
follows,\begin{equation}\langle {\cal B}_n \rangle \leq
1,\end{equation}which can be violated by quantum mechanics
\cite{MK}.

By convention, we adopt the notation $|0^n \rangle = |0 \cdots 0
\rangle$ and $|1^n \rangle = |1 \cdots 1 \rangle.$ For a pure
state $| \psi \rangle$ of $n$ qubits, the variance of ${\cal B}_n$
on $| \psi \rangle$ is defined by\begin{equation}\Delta \left (|
\psi \rangle, {\cal B}_n \right ) = \langle \psi |{\cal B}^2_n |
\psi \rangle - \langle \psi | {\cal B}_n | \psi
\rangle^2.\end{equation}Recall that when all the $\vec{a}_j$ and
$\vec{a}'_j$ are in the $x-y$ plane and $\vec{a}_j$'s are
distributed with angles $(j-1)(-1)^{n+1}2 \pi / (2n)$ with respect
to the $x$-axis and $\vec{a}_j \perp \vec{a}'_j,$ the associated
MK Bell operator has the following spectral decomposition
\cite{SG}\begin{equation}\hat{{\cal B}}_n = 2^{(n-1)/2}\left (
|\textrm{GHZ}_+\rangle \langle \textrm{GHZ}_+ | -
|\textrm{GHZ}_-\rangle \langle \textrm{GHZ}_- | \right
),\end{equation}where $|\textrm{GHZ}_{\pm} \rangle =
\frac{1}{\sqrt{2}}\left ( |0^n \rangle \pm |1^n \rangle \right )$
are GHZ's states \cite{GHZ}. Then, it is easy to check that
$\Delta \left (| 0^n \rangle, \hat{{\cal B}}_n \right ) =
2^{n-1}.$ Moreover, for every product state $| \psi \rangle = |
\psi_1 \rangle\cdot\cdot\cdot| \psi_n \rangle$ of $n$ qubits,
there is a local unitary transformation $U=U_1 \otimes
\cdot\cdot\cdot \otimes U_n$ such that $U | \psi \rangle = |0^n
\rangle.$ Hence,\begin{equation}\Delta \left (| \psi \rangle,
U^{\dagger} \hat{{\cal B}}_n U \right ) =
2^{n-1},\end{equation}where $U^{\dagger}$ denotes the adjoint
operator of $U.$ This leads the following

{\it Theorem}.-- A pure state $| \psi \rangle$ of $n$ qubits is
entangled if and only if\begin{equation}\Delta \left (| \psi
\rangle, U^{\dagger} \hat{{\cal B}}_n U \right ) <
2^{n-1},\end{equation}for every local unitary transformations $U$
such that $\langle 0^n| U | \psi \rangle$ and $\langle 1^n| U |
\psi \rangle$ are both real numbers.

Furthermore, $| \psi \rangle$ is entangled whenever there is one
$U$ satisfying Eq.(6) and solving the maximum
problem\begin{equation} \max \left ( \langle 0^n| U | \psi
\rangle^2 + \langle 1^n |U |\psi\rangle^2 \right
),\end{equation}where the maximum is taken over all local unitary
transformations so that $\langle 0^n| U | \psi \rangle$ and
$\langle 1^n| U | \psi \rangle$ are both nonnegative numbers.

{\it Proof}.--The sufficiency of {\it Theorem} is shown above. To
prove the necessity, suppose that there were ${\cal B}_n =
U^{\dagger} \hat{{\cal B}}_n U$ for some $U$ so that $\Delta \left
(| \psi \rangle, {\cal B}_n \right ) = 2^{n-1}$ and both $\langle
0^n| U | \psi \rangle$ and $\langle 1^n| U | \psi \rangle$ are
real numbers. Since $\| {\cal B}_n \| \leq 2^{(n-1)/2},$ it is
concluded that\begin{equation}{\cal B}^2_n| \psi \rangle = 2^{n-1}
| \psi \rangle,\end{equation}\begin{equation}\langle \psi | {\cal
B}_n | \psi \rangle =0.\end{equation}As proved in \cite{Chen},
Eq.(8) implies that$$ U | \psi \rangle = \alpha |0^n \rangle +
\beta |1^n \rangle.$$ Since $\alpha = \langle 0^n| U | \psi
\rangle$ and $\beta = \langle 1^n| U | \psi \rangle$ are both real
numbers, it is concluded from Eqs.(4) and (9) that either $\alpha$
or $\beta$ is zero and hence $| \psi \rangle$ is unentangled.

Furthermore, if $| \psi \rangle$ is a product state, then the
maximum value of Eq.(7) is one. Suppose that $\tilde{U}$ is a
solution of Eq.(7), then $\tilde{U} | \psi \rangle$ is a product
state of the form $\alpha |0^n \rangle + \beta |1^n
 \rangle$ with $\alpha, \beta$ being two nonnegative numbers
and hence must be equal to either $|0^n \rangle$ or $|1^n
\rangle,$ from which one concludes that $\Delta \left (| \psi
\rangle, \tilde{U}^{\dagger} \hat{{\cal B}}_n \tilde{U} \right ) =
2^{n-1}.$ This completes the proof.

By our criterion, it is easy to test the entanglement of the
generalized GHZ states\begin{equation}|\Psi (\varphi)\rangle =
\cos \varphi |0^n \rangle + \sin \varphi |1^n
\rangle\end{equation}with $0 \leq \varphi \leq \pi /4.$ Indeed,
since the identity $I=I_2 \otimes \cdot\cdot\cdot \otimes I_2$ is
a solution of the maximum problem Eq.(7), where $I_2$ is the
identity on one qubit, we obtain that\begin{equation}\Delta \left
(| \Psi (\varphi) \rangle, \hat{{\cal B}}_n \right ) = 2^{n-1}
\cos^2 2 \varphi.\end{equation}This exactly detects the
entanglement of $| \Psi (\varphi) \rangle,$ which, however, cannot
be tested by any Bell inequality of two dichotomic observables per
site for sufficiently small parameters $\varphi$ when $n>2,$ as
shown in \cite{ZBLW}.

Evidently, our criterion reduces to a quadratic program consisting
of maximizing a quadratic objective Eq.(7) under two linear
inequalities $\langle 0^n| U | \psi \rangle, \langle 1^n| U | \psi
\rangle \geq 0$ for all local unitary transformations of $n$
qubits. The complexity of solving this quadratic program will be
studied in the future. A similar argument using convex
optimization programs is presented in \cite{Program}.

Although our criterion is more powerful than the Bell inequalities
of two dichotomic observables per site for detecting the
pure-state entanglement of $n$ qubits, it cannot be used to test
quantum mechanics versus EPR's local realism from which the
original Bell inequality arose \cite{Local}. Indeed, as shown in
\cite{R}, the physical origins of EPR's local realism and quantum
entanglement are different. For a multipartite system which,
having interacted in past, are now spatially separated, EPR's
local realism means that elements of physical reality for one
subsystem should be independent on what are done with the others.
In contrary, quantum entanglement refers only to quantum
multipartite states, whether or not the individual subsystems are
spatially separated. The violation of Bell's inequalities assuming
EPR's local realism by suitable entangled states is therefore an
interesting but indirect consequence of quantum entanglement.

In summary, we have shown a Gisin-type theorem for $n$ qubits in
terms of variances of Mermin-Klyshko's Bell operators, which
cannot be straightforwardly generalized to all $n$-qubit systems
for the case of the Bell inequalities of two dichotomic
observables per site involving the mean values of observables
\cite{ZBLW}. Entanglement criteria based on inequalities for
variances of observables have been studied, mainly designed for
continuous variable systems \cite{V1}. Recently, it has been shown
that variances of observables can be detect entanglement also in
finite-dimensional systems \cite{V2}. Our result furthermore shed
considerable light on quantum entanglement in terms of variances
of the Bell operators of a special type. We expect that, similarly
to the Bell inequalities based on {\it mean values} of the
Bell-type operators, the {\it variances} of Bell-type operators
and its various ramifications will play an important role in
quantum entanglement.

This work was supported by the National Natural Science Foundation
of China under Grant No.10571176, the National Basic Research
Programme of China under Grant No.2001CB309309, and also funds
from Chinese Academy of Sciences.



\begin{thebibliography}{**}
\bibitem{Local}According to EPR's local realism, MK Bell operators
take only values $\pm 1.$ Then, their variances are less than or
equal to one and display that there is no contradiction betweem
quantum mechanics and EPR's local realism.
\end{thebibliography}
\end{document}